\documentclass[english,twocolumn,prl,showpacs]{revtex4-1}
\usepackage[T1]{fontenc}
\usepackage[latin9]{inputenc}
\usepackage{verbatim}
\usepackage{amsmath}
\usepackage{amssymb}
\usepackage{graphicx}
\usepackage{esint}
\usepackage{soul}

\makeatletter
\@ifundefined{textcolor}{}
{%
\definecolor{BLACK}{gray}{0}
\definecolor{WHITE}{gray}{1}
\definecolor{RED}{rgb}{1,0,0}
\definecolor{GREEN}{rgb}{0,1,0}
\definecolor{BLUE}{rgb}{0,0,1}
\definecolor{CYAN}{cmyk}{1,0,0,0}
\definecolor{MAGENTA}{cmyk}{0,1,0,0}
\definecolor{YELLOW}{cmyk}{0,0,1,0}
}
\usepackage{hyperref}
\usepackage{xcolor}
\hypersetup{
    colorlinks,
    citecolor=blue,
    filecolor=blue,
    linkcolor=blue,
    urlcolor=blue,
    pdfstartview=FitH
}
\newcommand{\co}[1]{#1}

\makeatother
\usepackage{babel}
\begin{document}

\title{Non-Majorana subgap states in two-dimensional topological insulator-superconductor hybrid structures}

\author{Hoi-Yin Hui}

\affiliation{Department of Physics, Condensed Matter Theory Center and Joint Quantum
Institute, University of Maryland, College Park, Maryland 20742-4111,
USA}

\author{Jay D. Sau}

\affiliation{Department of Physics, Condensed Matter Theory Center and Joint Quantum
Institute, University of Maryland, College Park, Maryland 20742-4111,
USA}

\author{S. Das Sarma}

\affiliation{Department of Physics, Condensed Matter Theory Center and Joint Quantum
Institute, University of Maryland, College Park, Maryland 20742-4111,
USA}

\date{\today}
\begin{abstract}
Contrary to the widespread belief that Majorana zero-energy modes,
existing as bound edge states in 2D topological insulator (TI)-superconductor
(SC) hybrid structures, are unaffected by non-magnetic static disorder
by virtue of \co{time-reversal symmetry}, we show that such a protection against
disorder does not exist in realistic multi-channel TI/SC/ferromagnetic
insulator (FI) sandwich structures of experimental relevance since
the time-reversal symmetry is explicitly broken locally at the SC/FI
interface where the end Majorana mode (MM) resides. We find that although
the MM itself and the \emph{bulk} topological superconducting phase
inside the TI are indeed universally protected against disorder, extra fermionic
subgap states are generically introduced at the TI edge due to the
presence of the FI/SC interface as long as multiple edge channels
are occupied. We discuss the important implications of our finding for the detection
and manipulation of the edge MM in TI/SC/FI experimental
systems of current interest.

\end{abstract}


\maketitle

\begin{figure}
\begin{centering}
\includegraphics[width=0.85\columnwidth]{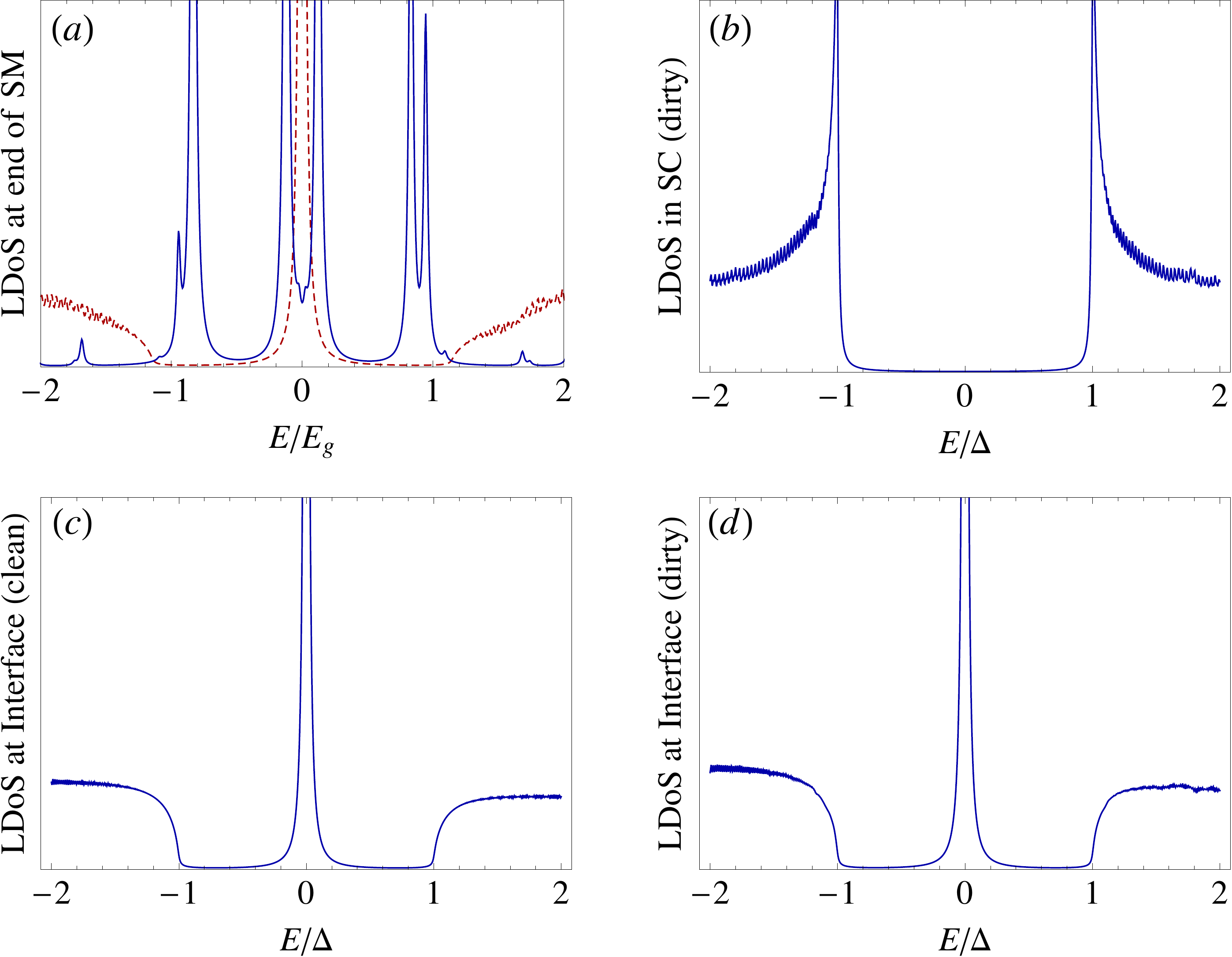}
\par\end{centering}

\caption{(a) LDOS at the end of a SM nanowire, where
the dashed red line shows the result for a clean wire while the solid
blue line is for a disordered wire.
The parameters used for the BdG Hamiltonian $H=(\frac{p^2}{2m}+\alpha p\sigma_y-\mu)\tau_z+B\sigma_z+\Delta\tau_x$ are: $\mu=0$, $B=2\Delta$, $m\alpha^2=2\Delta$.
(b) LDOS deep in the SC region of the TI edge, with disorder in SC region.
(c) LDOS at the SC/FI interface on a clean TI. (d) LDOS
at the SC/FI interface on a TI edge with disorder in SC region. \label{fig:1-channel}}

\end{figure}

Current interest \cite{Beenakker2013Search,Alicea2012New,Leijnse2012Introduction,Stanescu2013Majorana}
in the search for solid-state Majorana modes (MMs), i.e. zero-energy
Majorana bound states inside a superconducting gap, has increased sharply
following a series of experimental reports \cite{Mourik2012Signatures,Deng2012Anomalous,Das2012Zero-bias,Churchill2013Superconductor-nanowire,Finck2013Anomalous,Rokhinson2012fractional}
claiming empirical evidence for the existence of MMs in semiconductor
(SM) nanowires in proximity to superconductors (SCs) in the presence
of an external magnetic field precisely as predicted in earlier theoretical
publications \cite{Lutchyn2010Majorana,Sau2010Non-Abelian,Oreg2010Helical}.
Very recent MM-motivated experimental interest \cite{Williams2012Unconventional,*Knez2011Evidence,*Knez2012Andreev,*Qu2012Strong,*Du2013Observation,*Oostinga2013Josephson,*Kurter2013Evidence,*Orlyanchik2013Signature,*Xu2013Majorana,*Hart2013Induced,*Pribiag2014Edge-mode} 
has focused also on
2D (i.e. the so-called quantum spin Hall system) and 3D topological
insulators (TIs) in proximity to ordinary $s$-wave SCs, where, in fact,
the first theoretical proposals for the existence of MMs were originally
made by Fu and Kane \cite{Fu2008Superconducting,Fu2009Josephson,*Fu2009Probing},
and then followed up by others \cite{Akhmerov2009Electrically,*Linder2010Unconventional,*Xu2010Fractionalization,*Cook2011Majorana,*Mi2013Proposal}.
One substantive advantage of the TI-based MM proposals over the SM-based
proposals is that the explicit presence of time-reversal symmetry
(TRS) in the TI/SC system makes the bulk topological
SC phase in the TI and the associated MMs immune to static non-magnetic
disorder arising from impurities and defects invariably present
in the environment. By contrast, the SM/SC hybrid topological structures
hosting MMs are unprotected from non-magnetic disorder in the environment {[}see Fig.~\ref{fig:1-channel}(a) for an illustrative example in which the MM is \co{destroyed} by strong disorder{]}
since time-reversal invariance must be explicitly broken \cite{Sau2010Generic}
in the SM/SC sandwich structures in order to produce the MM-carrying
topological phase. We emphasize, however, that both the TI/SC and the SM/SC
systems are protected from elastic disorder residing in the bulk
superconductor itself \cite{Stanescu2011Majorana} which can have
no effect on the proximity-induced topological superconductivity in
the TI or the SM material -- the protection (or not) for the TI (or
SM) system that we are discussing here is specifically from the disorder
residing at the interface or inside the bulk TI (or SM) material.

\begin{figure}
\begin{centering}
\includegraphics[width=0.7\columnwidth]{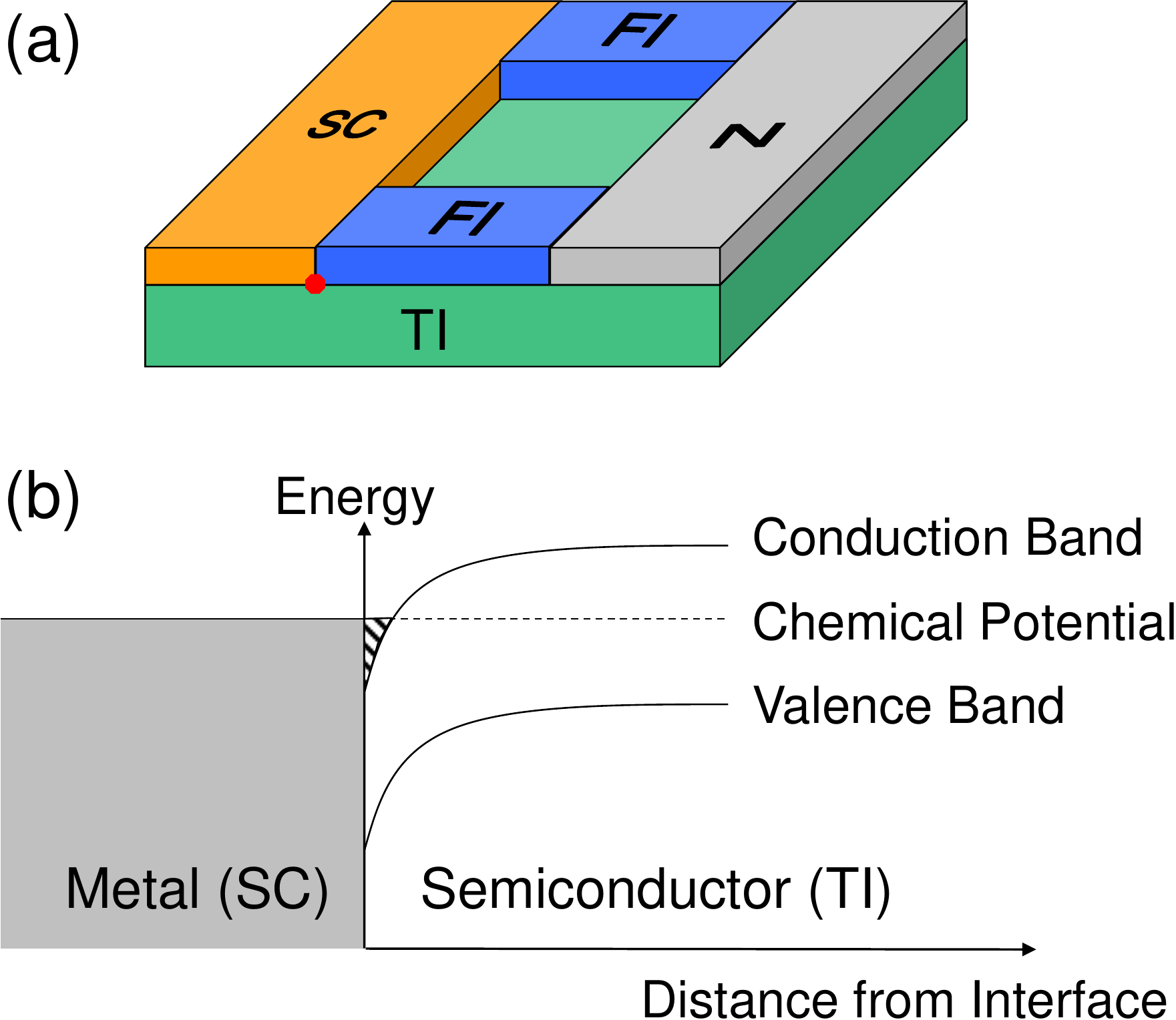}
\par\end{centering}

\caption{(a) Schematic picture of the devices: a TI in contact with SCs and
an FI . The position of MM is marked with a red dot. The normal metal N acts
as an external lead and the FI acts as a tunnel barrier. The MM
is then detected as a zero-bias conductance peak.  (b) At the interface
between the metallic SC and the narrow-gap semiconductor (i.e. the
TI), in general band-bending effect at the interface leads to extra
edge channels (hashed region). \label{fig:device}}
\end{figure}

Although there have been many theoretical studies \cite{Akhmerov2011Quantized,Pientka2012Enhanced,*Sau2012Experimental,*Brouwer2011Topological,*Neven2013Quasiclassical,*Adagideli2013Inducing}
of disorder effects on the MMs and the topological superconductivity
in the SM/SC nanowire systems of experimental relevance \cite{Mourik2012Signatures,Deng2012Anomalous,Das2012Zero-bias,Churchill2013Superconductor-nanowire,Finck2013Anomalous,Rokhinson2012fractional},
including even some suggestions that the experimental observations
of the zero bias peak \cite{Mourik2012Signatures,Deng2012Anomalous,Das2012Zero-bias,Churchill2013Superconductor-nanowire,Finck2013Anomalous}
in the InSb and InAs semiconductor nanowires in the SM/SC hybrid structures,
thought to be arising from the zero-energy bound MMs localized at
the wire ends as predicted theoretically \cite{Lutchyn2010Majorana,Sau2010Non-Abelian,Oreg2010Helical},
are in fact direct manifestations of anti-localization effects induced
by static disorder \cite{Pikulin2012zero-voltage}, there has been no theoretical
analysis of disorder effects in the TI/SC hybrid structures except
for the work in Ref.~\cite{Potter2011Engineering} which concluded
that TI/SC topological systems are completely protected from \emph{all} elastic
disorder effects by virtue of Anderson's theorem \cite{Anderson1959Theory} due to TRS. A recent study, however, concluded that the induced $p$-wave superconductivity
may in general be suppressed by disorder in a 3D TI/SC structure \cite{Tkachov2013Suppression}.

In the current work, we consider the experimental TI/SC
structure for the existence of bound MMs where a ferromagnetic insulator (FI) must
be deposited in order to localize the MMs at the system edge. The realistic structure {[}Fig.~\ref{fig:device}(a){]},
first proposed by Fu and Kane in this context \cite{Fu2008Superconducting},
involves the 2D TI with SC and FI layers deposited on top of
it, referred to as the STIM (superconductor-TI-magnet) system in Ref.~\cite{Fu2008Superconducting}.
The FI layer in the STIM breaks TRS which is essential for localizing
spatially separated topologically non-Abelian MMs.
The absolute necessity of TRS breaking for creating isolated defect-bound
MMs is a well-established theorem \cite{Leijnse2012Introduction}.

In this scenario, the immunity of the local SC gap in the TI/SC/FI
structure against disorder is subtle. Although Anderson's theorem
guarantees that no impurities can degrade the gap deep in the SC region \co{where TRS
applies}, there is no corresponding
argument near the SC/FI interface where the FI provides an explicit TRS-breaking
mechanism \co{making the theorem inapplicable}. This means that although the MMs separated by the SC do
not hybridize, extra fermionic subgap states could potentially appear locally
at the SC/FI interface in the TI/SC/FI hybrid structure. 
Our current work is the first theoretical exploration of the existence (or not)
of such subgap states in STIM systems.

In spite of the explicit TRS-breaking, we find that disorder in the SC region of
the TI edge is unable to generate any extra localized fermionic subgap states
(i.e. in addition to the zero-energy MM itself) near
the SC/FI interface {[}see Fig.~\ref{fig:1-channel}(d){]}, provided that we limit our
attention to only single-channel TI edges. Thus, the TI/SC/FI hybrid
system is indeed immune to all disorder provided there is only a single
active edge channel in the system. However, disorder-induced potential
fluctuations near the edge are expected to produce bound states in addition to the 1D edge state.
Such extra states or puddles, which have been proposed to explain the temperature dependence
of the TI edge conductance \cite{Vayrynen2013Helical}, can be modeled
using a multi-channel TI edge. In fact, in realistic structures, we
expect multi-channel 1D edges in the generic 2D TI system due to, for example,
band-bending effects which are ubiquitous near semiconductor surfaces.


We investigate a broad class of TI Hamiltonians by considering
a multi-channel TI edge. In addition to disorder, the extra 1D edge channels
could be induced from an intrinsically higher chemical potential near
the surface \cite{Bianchi2010Coexistence,Bahramy2012Emergent}. Also, since the
2D TI is commonly constructed from semiconductors with small band gaps (e.g.
HgCdTe or InAs/GaSb) \cite{Bernevig2006Quantum,Koenig2007Quantum},
a proximate metallic SC would unavoidably induce extra edge channels due to
band-bending \cite{Kittel1986Introduction} {[}see Fig.~\ref{fig:device}(b){]}.

We consider the following Hamiltonian which models a multi-channel
TI edge in proximity to SC/FI:

\begin{equation}
\begin{split}H= & \int dx\left\{ v\sum_{\alpha ss'}\left(\boldsymbol{p}_{\alpha}\cdot\boldsymbol{\sigma}_{ss'}\right)\psi_{\alpha s}^{\dagger}(x)\left(-i\partial_{x}\right)\psi_{\alpha s'}(x)\right.\\
 & -\sum_{\alpha\beta s}\mu_{\alpha\beta}(x)\psi_{\alpha s}^{\dagger}(x)\psi_{\beta s}(x)\\
 & -\sum_{\alpha\beta iss'}\tilde{\mu}_{\alpha\beta}^{j}(x)\imath\sigma_{ss'}^{j}\psi_{\alpha s}^{\dagger}(x)\psi_{\beta s'}(x)\\
 & +\sum_{\alpha}\Delta_{\alpha}(x)\left[\psi_{\alpha\uparrow}(x)\psi_{\alpha\downarrow}(x)+{\rm hc}\right]\\
 & \left.+\sum_{\alpha ss'}B_{\alpha}(x)\sigma_{ss'}^{x}\psi_{\alpha s}^{\dagger}(x)\psi_{\alpha s'}(x)\right\} 
\end{split}
\label{eq:H}
\end{equation}
Here, $\psi_{\alpha s}^{\dagger}(x)$ creates an electron in the $\alpha^{{\rm th}}$
channel with spin $s=\uparrow,\downarrow$ at position $x$, where each "channel" has a Kramers pair
of bands. \co{Note that the nature of TI edge requires the number of channels $N_{\rm ch}$ to be odd.}
$\boldsymbol{\sigma}=\left(\sigma^{x},\sigma^{y},\sigma^{z}\right)$
are the three Pauli spin matrices, and $\boldsymbol{p}_{\alpha}$
is the polarization of the $\alpha^{{\rm th}}$ channel. The $N_{{\rm ch}}\times N_{{\rm ch}}$
real symmetric matrix $\mu_{\alpha\beta}(x)$ contains both the chemical
potential of each channel (diagonal entries) and inter-channel elastic
scattering (off-diagonal entries), while the anti-symmetric matrix
$\tilde{\mu}_{\alpha\beta}^{j}(x)$ are the coefficients for inter-channel
spin-orbit scatterings, which still respect TRS. 
The proximate SC and FI induce the local pairing potential $\Delta(x)$
and the Zeeman term $B(x)$ respectively. To model an SC/FI interface,
we choose $\Delta(x)=\Delta\theta(x)$ and $B(x)=B\theta(-x)$ where $B=3\Delta$, restricting
the analysis to an idealized case where there are no spatial fluctuations
of $\Delta$ and $B$ on the edge \cite{Takei2013Soft} \co{and no penetration of 
$\Delta$ and $B$ to the FI and SC side respectively, since this would degrade the spectral gap at the 
SC/FI interface in a trivial way and obscure the main finding of this work}. We also
assume that there are no inter-channel pairings or Zeeman gaps.

Static charge impurities and spin-orbit impurities are included through spatial variations
in $\mu(x)$ and $\tilde{\mu}(x)$ respectively. For the results below in which disorder is included, 
the mean free path $l$ is estimated to be $k_Fl\approx3$. If the disorder in either $\mu$ or $\tilde{\mu}$
is removed, the phase space of scattering is reduced, but similar results would still apply
provided that the mean free path remains the same \co{by tuning up $\tilde{\mu}$ or $\mu$ respectively}.
 Note that disorder is introduced in the SC region only
since its effect on the FI region can in principle be offset by a sufficiently strong Zeeman term.


Before presenting the results based on numerical simulations, we first
analyze the problem with a scattering matrix approach \cite{Akhmerov2011Quantized,Wimmer2011Quantum},
the details of which are presented in the Appendix.
We treat the SC/FI interface as an SC-N-FI system where the N region
has a finite but vanishingly small width. Then, the localized modes
at the interface are found by the condition $\det\left(1-\tilde{R}R\right)=0$
where $R(E)$ and $\tilde{R}(E)$ are respectively the reflection
matrices at the SC-N and N-FI interfaces. At zero energy, $R$ is
constrained by unitarity, time-reversal symmetry, and particle-hole
symmetry, while $\tilde{R}$ is constrained by unitarity and U(1)
symmetry. From these constraints, we can show that $1\leq N_{0}\leq N_{{\rm ch}}$,
where $N_{0}$ is the multiplicity of the unity eigenvalue of $\tilde{R}R$,
which is equal to the number of localized zero-energy modes.

The exact value of $N_{0}$ depends on the details of the Hamiltonian,
but one can always fine-tune the Hamiltonian near the SC/FI interface
by local disorder respecting the symmetries, such that all energies
of the localized modes reach zero. Therefore the constraint derived
above implies that the number of localized states is equal to the
number of channels in the TI edge, and among them there is always
a zero-energy mode, which is the MM.

\co{We have therefore shown that the gap protecting the MM is indeed robust in the single-channel case.
This is not a consequence of Anderson's theorem as commonly believed, but is due to the various symmetries
present in the system as elucidated by the above scattering matrix argument.
Note that in the multi-channel case, although subgap states exist,
the MM is still pinned at zero energy.}

In the following we numerically
study how the number of subgap states $N_{0}$ depends on the number
of channels and other details of the interface.

\begin{figure}
\begin{centering}
\includegraphics[width=0.85\columnwidth]{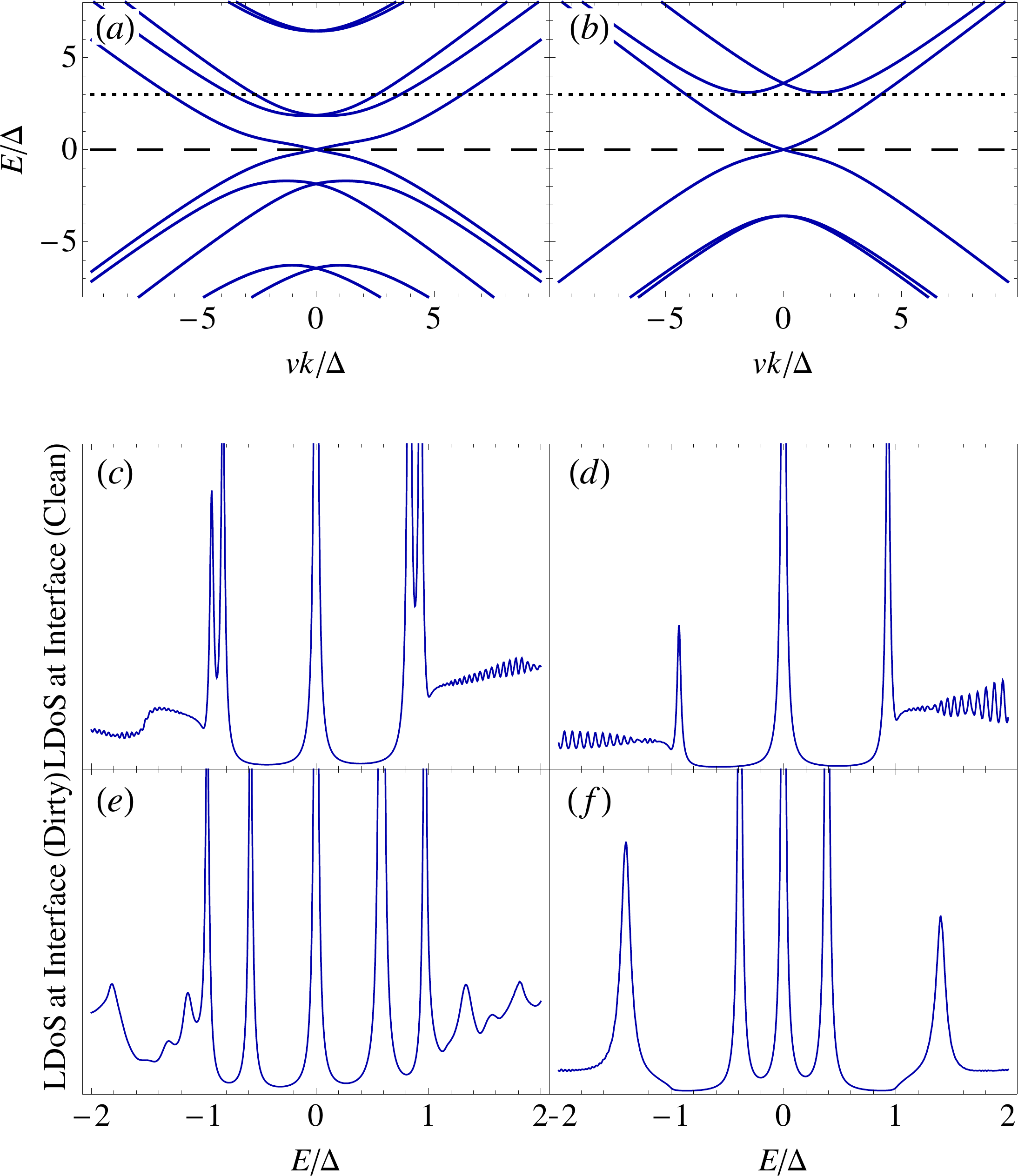}
\par\end{centering}

\caption{(a,b) Band structures of respectively a 5-channel and 3-channel TI
edge. Here the parameters of Eq.~(\ref{eq:H}) is chosen as: $\mu={\rm diag}\left(2,-2,1,-1,0\right)\Delta$,
$\tilde{\mu}^{x}=\tilde{\mu}^{y}=0$, $\tilde{\mu}_{\alpha\gtrless\beta}^{z}=\pm2\Delta$,
$\boldsymbol{p}_{\alpha}=\left(+,-,+,-,+\right)\hat{z}$. The Fermi
level in the FI region ($E_{F}=0$) is shown in dashed line, while
that in the SC region ($E_{F}=3\Delta$) is shown in dotted line.
(c,d) Their corresponding LDOS of at the SC/FI interface, without
disorder. (e,f) Their corresponding LDOS at the SC/FI interface, with
a single realization of non-magnetic disorder.\label{fig:multichannel}}
\end{figure}

We start by considering the simplest case of a TI/SC/FI interface
where the TI has a single channel. The local density of states (LDOS)
\cite{Gennes1989Superconductivty,*Gygi1990Electronic} at the SC/FI interface
with or without disorder is plotted respectively in Fig.~\ref{fig:1-channel}(c,d).
It shows that the LDOS in the subgap regime $\left(E<\Delta\right)$
is not affected by disorder in the single-channel case. Fig.~\ref{fig:1-channel}(b) shows that
the gap deep in the SC region is also completely unaffected. This is
consistent with previous results \cite{Potter2011Engineering} and our scattering matrix analysis
for the single-channel case. This simplification, however, disappears as soon as
the system has multichannel edge states as is likely in realistic samples (Fig.~\ref{fig:multichannel}).

To understand the interplay of disorder and multiple channels we consider
5-channel and 3-channel TI edge models. The parameters of Eq.~(\ref{eq:H})
are so chosen so that the additional channels cross the Fermi level
in the SC region. It is assumed that the FI region is gated so as to place
the chemical potential in the gap of the additional  bands as well
{[}see Fig.~\ref{fig:multichannel}(a,b){]}. This is necessary to
obtain localized MMs at the interface. The LDOS at the interface for
several cases are plotted in Fig.~\ref{fig:multichannel}(c,d,e,f).
We typically find that each channel leads
to a subgap state. The LDOS in Figs.~\ref{fig:multichannel}(c,d)
show that for the interface parameters chosen, the interface states
are close to the edge of the gap away from the interface. On the other
hand, introducing disorder leads to the results in Figs.~\ref{fig:multichannel}(e,f), where
one sees interface states that are bound deep inside the bulk gap.
The details of the SC/FI interface, such as the chemical potential
change at the interface, can lead to scattering very similar to disorder effects.
Therefore, depending on the details of the
interface it is possible even for an interface without any explicit quenched disorder to
have subgap states in the middle of the gap. The results plotted in
Figs.~\ref{fig:multichannel}(c,d) correspond to a smooth interface
where such scattering is absent and therefore do not contain deep
subgap states. In contrast to the interface LDOS, the LDOS in the
SC region, away from the SC/FI interface, does not show any subgap states
similar to the single channel case. \co{This could be understood as a consequence
of TRS, by which the application of Anderson's theorem forbids such subgap states.}

Comparing the results of the disordered SC/FI interface in the single
channel case (Fig.~\ref{fig:1-channel}) and the multi-channel
case (Fig~\ref{fig:multichannel}) it is clear that the absence
of subgap states in the single channel case is not simply a result
of Anderson's theorem (i.e. TRS). As mentioned in the introduction,
since TRS is locally broken at the interface, it only protects
the bulk, i.e. the region away from the SC/FI interface, from subgap
states. On the other hand, the scattering-matrix interpretation 
implies that the presence of a single subgap state in the
single channel case is a result of the properties of the scattering
matrix near zero energies. Time-reversal invariance does of course
play a role in determining the properties of the scattering matrix.
The scattering matrix picture also allows for the presence of extra
subgap states in the multi-channel case as seen in Fig.~\ref{fig:multichannel}.

\begin{figure}
\begin{centering}
\includegraphics[width=1\columnwidth]{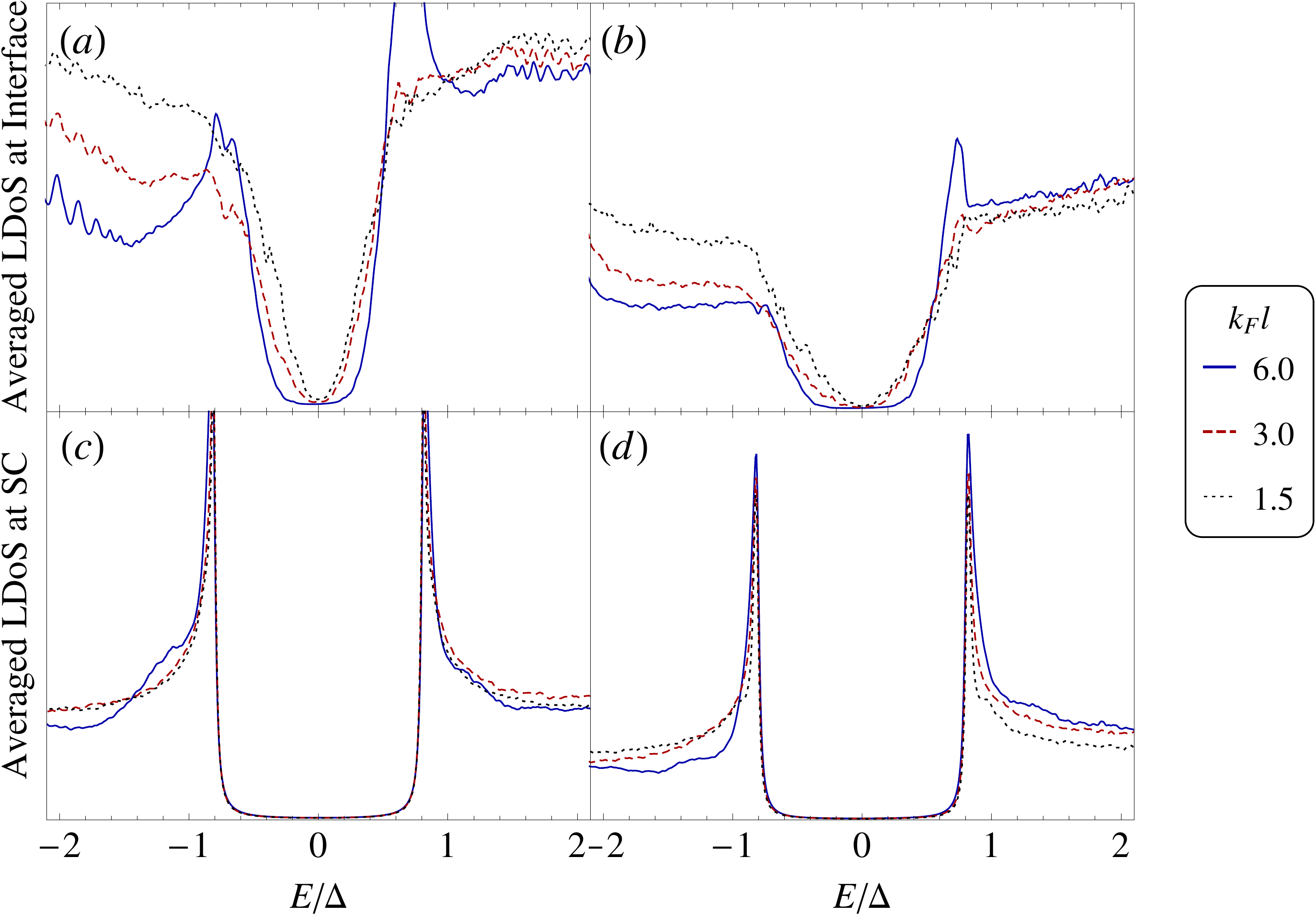}
\par\end{centering}

\caption{(a,b) LDOS at the SC/FI interface, averaged over 5000 realizations
of non-magnetic disorder, for a 5-channel and 3-channel TI edge respectively.
The three curves show the results for different strengths of disorder,
and the MMs have been removed for clarity (c,d) Their corresponding
averaged LDOS deep in the SC region. \label{fig:multichannelAv}}
\end{figure}

The results in Fig.~\ref{fig:multichannel} are for a typical (rather than fine-tuned)
structure of the interface (which is determined by the potentials in
the various channels) and also for typical disorder configurations
and should thus be regarded as typical results which should be qualitatively valid generically.
While these results clearly establish the generic existence of fermionic subgap states
in a multi-channel TI with disorder, the experimentally relevant
question might be the likelihood of the occurrence of these subgap states 
with a random disorder potential. To address this question
we calculate the disorder-averaged LDOS
at the interface (Fig.~\ref{fig:multichannelAv}). As seen from Fig.~\ref{fig:multichannelAv}(a,b),
the superconducting gap at the SC/FI interface is not protected against
disorder and is reduced by increasing disorder strength. Interestingly,
the disorder-averaged LDOS for strong disorder is found
to vanish at zero energy. This suggests that level repulsion from
the MM prevents the extra subgap states from approaching precise
zero-energy which is consistent with conclusions from random matrix theory \cite{Ivanov2002supersymmetric,*Bagrets2012Class,*Ioselevich2012Majorana}.
In Fig.~\ref{fig:multichannelAv}(c,d) we show
the disorder-averaged gap in the TI/SC edge away from the SC/FI interface.
In contrast to the gap near SC/FI interface, we find that the TI/SC gap is immune
to disorder \co{as expected since TRS is respected there and Anderson's theorem applies}. 
It is instructive to note that the LDOS at the interface {[}Fig.~\ref{fig:multichannelAv}(a),(d){]}
shows precisely the same soft gap behavior widely observed in the
SM/SC nanowire hybrid systems \cite{Takei2013Soft,Stanescu2014Soft}.

The additional subgap states may have consequences for detecting and
manipulating MMs at finite temperatures.
One of the simplest signatures of an MM is the zero-bias tunneling conductance
peak. The voltage resolution
of a tunneling conductance measurement, which directly probes the LDOS,
is limited by the finite tunneling rate and temperature.
Low-energy subgap states as in Fig.~\ref{fig:multichannel} and Fig.~\ref{fig:multichannelAv}
would contribute to tunneling if
the energy of the subgap states is lower than either the temperature
or the tunneling-induced broadening.  However, since there is
a repulsion of these states from zero energy {[}Fig.~\ref{fig:multichannelAv}(a,b){]},
the likelihood for the states to influence the zero-bias conductance
is small unless the temperature is high and the number of channels
is large. Thus, the zero-bias effective MM tunneling conductance peak may survive the existence of fermionic subgap states.
For utilizing the MM for topological quantum computation,
it is known that extra localized states do not affect
the phase of manipulation \cite{Akhmerov2010Topological}. However,
the additional subgap levels could have an influence for the readout
schemes that rely on measuring single-particle spectra \cite{Flensberg2011Non-Abelian},
because this requires eliminating the TR-breaking region to hybridize
the two MMs to finite energies. In spite of this, the TI/SC structure
has an advantage in regards to robustness against disorder, since
disorder cannot degrade the SC gap deep in the SC region thus hybridizing
the two MMs at the ends. Also, the potential-fluctuation-induced subgap states
show a repulsion from zero energy, without any complications arising
from weak antilocalization \cite{Pikulin2012zero-voltage,Sau2013Density}.

Experimentally, the conductance of a TI edge was measured as $2e^{2}/h$
\cite{Koenig2007Quantum}, which might appear to contradict our assumption
of multiple edge channels. However, the simple relation $N_{{\rm ch}}=\frac{G}{(2e^{2}/h)}$
between conductance $G$ and the number of channels $N_{{\rm ch}}$
is valid in the clean limit only. With disorder, the conductance should
be analyzed using random matrix theory of the edge transmission matrix \cite{Beenakker1997Random-matrix}.
Since the multi-channel TI edge belongs to the symplectic class with
odd number of channels, only one channel remains delocalized \cite{Takane2004Conductance}.
A conductance measurement with a length scale greater than the localization
length would then produce $G=2e^{2}/h$ even if many channels are
present. Therefore, the conductance measurement alone cannot rule out
the presence of multiple channels.

To conclude, we have studied the effects of multiple channels and
disorder near an SC/FI interface on the edge of a 2D TI in the context of MM in the system.
We find that a number of localized
states, equal to the number of channels, appears at the SC/FI interface.
One of these states is the zero-energy MM while the energies of the
other states depend on the specific details of the system. 
Adding disorder in the SC region leads to a distribution in the energies
of these extra localized states, potentially reaching the subgap regime.
However their effect is less detrimental than those in the SM/SC nanowire
structures since they are repelled from zero energy and are localized
at the boundary.

This work is supported by JQI-NSF-PFC and Microsoft Q.

\appendix
\section{Scattering Matrix Approach}

In this appendix we present the details of the derivation
of zero-energy modes by scattering matrices.

Consider an SC-N-FI problem, such that the interface between SC and
FI is expanded to a finite width. A solution in N is
\begin{equation}
\psi^{\dagger}=u_{+}\psi_{+}^{\dagger}+u_{-}\psi_{-}^{\dagger}+v_{-}\psi_{-}+v_{+}\psi_{+},
\end{equation}
in which the multi-component $\psi_{\pm}$ are Kramer's pairs and
the subscripts $\pm$ represents right/left moving modes. The corresponding
particle-hole-conjugated and time-reversed solutions are:
\begin{eqnarray}
\mathcal{C}\psi^{\dagger}\mathcal{C}^{-1} & = & v_{+}^{*}\psi_{+}^{\dagger}+v_{L}^{*}\psi_{-}^{\dagger}+u_{-}^{*}\psi_{-}+u_{+}^{*}\psi_{+}\\
\mathcal{T}\psi^{\dagger}\mathcal{T}^{-1} & = & -u_{-}^{*}\psi_{+}^{\dagger}+u_{+}^{*}\psi_{-}^{\dagger}+v_{+}^{*}\psi_{-}-v_{-}^{*}\psi_{+}
\end{eqnarray}

Reflection matrices $R$ relates the coefficients of $\left(\psi_{+}^{\dagger},\psi_{+}\right)$
to those of $\left(\psi_{-}^{\dagger},\psi_{-}\right)$. At the left
(SC/N) interface, the particle-hole symmetry and time-reversal symmetry
are both respected, giving three equations for $R$:

\begin{eqnarray}
\left(\begin{array}{c}
u_{+}\\
v_{+}
\end{array}\right) & = & R\left(\begin{array}{c}
u_{-}\\
v_{-}
\end{array}\right)\\
\left(\begin{array}{c}
v_{+}^{*}\\
u_{+}^{*}
\end{array}\right) & = & R\left(\begin{array}{c}
v_{-}^{*}\\
u_{-}^{*}
\end{array}\right)\\
\left(\begin{array}{c}
-u_{-}^{*}\\
-v_{-}^{*}
\end{array}\right) & = & R\left(\begin{array}{c}
u_{+}^{*}\\
v_{+}^{*}
\end{array}\right),
\end{eqnarray}
from which one can derive

\begin{equation}
R=\tau_{x}R^{*}\tau_{x}=-R^{T}.
\end{equation}
Together with the unitarity condition $R^{\dagger}R=1$, the value
of $\det R$ could be computed. First note that the constraint $R=\tau_{x}R^{*}\tau_{x}$
requires $R$ to take the form $R=\left(\begin{array}{cc}
r_{ee} & r_{eh}\\
r_{eh}^{*} & r_{ee}^{*}
\end{array}\right)$, while the condition $R=-R^{T}$ implies that $r_{ee}$ is antisymmetric, and, since its dimension is odd, it is singular.
Since $R$ itself is non-singular (by the unitarity condition), this implies that $r_{eh}$ is invertible.
The remaining conditions $R^{\dagger}R=1$ implies
\begin{eqnarray}
r_{eh}^{\dagger}r_{ee}+r_{ee}^{T}r_{eh}^{*}=0 & \Rightarrow & r_{ee}r_{eh}^{*-1}=-r_{eh}^{\dagger-1}r_{ee}^{T}\label{eq:zero}\\
r_{eh}^{\dagger}r_{eh}+r_{ee}^{T}r_{ee}^{*}=1 & \Rightarrow & r_{ee}^{T}r_{ee}^{*}=1-r_{eh}^{\dagger}r_{eh}\label{eq:one}\\
\end{eqnarray}
Now we evaluate $\det R$:
\begin{eqnarray}
\det R & = & \left(-1\right)^{N_{{\rm ch}}}\det\left(\begin{array}{cc}
r_{eh} & r_{ee}\\
r_{ee}^{*} & r_{eh}^{*}
\end{array}\right)\nonumber \\
 & = & \left(-1\right)^{N_{{\rm ch}}}\det\left(r_{eh}^{*}r_{eh}-r_{eh}^{*}r_{ee}r_{eh}^{*-1}r_{ee}^{*}\right)\nonumber \\
 & = & -\det\left(r_{eh}^{*}r_{eh}^{\dagger-1}\right) \\
 & = & -1\label{eq:detR}
\end{eqnarray}
where we have utilized the fact that $r_{eh}$ is invertible and Eqs.~(\ref{eq:zero}, \ref{eq:one}) are 
consecutively used in the intermediate steps. $N_{{\rm ch}}=\dim R/2$ is the number of channels
in the model, which is restricted to be odd due to the nature of the topological insulator edge.

At the right (N/FI) interface we have no Andreev reflection or time-reversal
symmetry, constraining the form of $\tilde{R}$ to be $\tilde{R}=\left(\begin{array}{cc}
\tilde{r}_{ee} & 0\\
0 & \tilde{r}_{ee}^{*}
\end{array}\right)$. By the unitarity of $\tilde{R}$, we have $\det\tilde{R}=1$.

The zero-energy modes are found by solving $\det\left(1-\tilde{R}R\right)=0$,
which implies that the multiplicity of the eigenvalue $-1$ of $-\tilde{R}R$
gives the number of zero-energy modes. We now prove that $-\tilde{R}R$\emph{
}must have at least one eigenvalue being $-1$\emph{.} To this end
we rotate $R$ to the Majorana basis via $\Omega=\frac{1}{\sqrt{2}}\left(\begin{array}{cc}
1 & 1\\
-i & i
\end{array}\right)$:
\begin{equation}
\Omega R\Omega^{\dagger}=2\left[\begin{array}{cc}
{\rm Re}\left(r_{ee}+r_{eh}\right) & -{\rm Im}\left(r_{ee}-r_{eh}\right)\\
{\rm Im}\left(r_{ee}-r_{eh}\right) & {\rm Re}\left(r_{ee}-r_{eh}\right)
\end{array}\right]\label{eq:MB}
\end{equation}
This matrix is real and, because it is unitary, it is also orthogonal.
Similarly for $\tilde{R}$ and hence $-\tilde{R}R$. The eigenvalues
of an orthogonal matrix can only be $1$, $-1$, or pairs of conjugate
$e^{\pm i\phi}$. Since $\det\left(-\tilde{R}R\right)=-1$, $-\tilde{R}R$
must have \emph{at least} one eigenvalue being $-1$. On the other
hand, because there are no other constraints on the problem, there
could be \emph{at most} $N_{{\rm ch}}$ eigenvalues being $-1$, and
hence, at most $N_{{\rm ch}}$ zero modes.

\vfill{}

\bibliographystyle{apsrev4-1}
\bibliography{TIdisorder}


\end{document}